\documentstyle[11pt]{article}
\newcommand{\be}{\begin{equation}}
\newcommand{\ee}{\end{equation}}
\newcommand{\ba}{\begin{eqnarray}}
\newcommand{\ea}{\end{eqnarray}}
\newcommand{\la}{\langle}
\newcommand{\ra}{\rangle}
\newcommand{\Mn}{ M_{\mbox{\tiny N}}}
\newcommand{\trF}{{\rm tr}_{\mbox{\tiny F}}}
\newcommand{\di}{ {\rm d} }
\newcommand{\btau}{{{\mbox{\boldmath$\tau$}}}}
\newcommand{\bgam}{{{\mbox{\boldmath$\gamma$}}}}
\newcommand{\bDelta}{{{\mbox{\boldmath$\Delta$}}}}
\newcommand{\bnabla}{{{\mbox{\boldmath$\nabla$}}}}
\newcommand{\ph}{{\phantom{\Biggl|}}}
\newcommand{\fslash}[1] {{\not\! #1\,}}
\newcommand{\binomial}[2]{\left(\mbox{$\!\!{\renewcommand{\arraystretch}{0.5}
  	\begin{array}{c} #1\\ #2\end{array}}\!\!$}\right)}
\newcommand{\singlesum}[1]{\sum_{\renewcommand{\arraystretch}{0.3}
  \begin{array}{l}\scriptscriptstyle{#1}\end{array}}}
\newcommand{\singlesumUp}[2]{\!\!\!\!\sum_{\renewcommand{\arraystretch}{0.3}
  \begin{array}{l}\scriptscriptstyle #1\end{array}}^{\scriptscriptstyle #2}
  \!\!}
\newcommand{\doublesumUp}[3]{\sum_{\renewcommand{\arraystretch}{0.6}
  \begin{array}{c}\scriptscriptstyle #1\\ \scriptscriptstyle #2\end{array}}
  ^{\scriptscriptstyle{#3}}}
\newcommand{\doublelim}[2]{{\lim_{\renewcommand{\arraystretch}{0.5}
  \begin{array}{l}\scriptscriptstyle #1\\ \scriptscriptstyle #2\end{array}}
  \!\!}}
\newcommand{\triplelim}[3]{{\lim_{\renewcommand{\arraystretch}{0.5}
  \begin{array}{c}\scriptscriptstyle {#1} \\ \scriptscriptstyle {#2} \\
  \scriptscriptstyle {#3}\end{array}} \!\!}}

\begin{document}
\title{	Polynomiality of unpolarized off-forward distribution functions \\
	and the \boldmath $D$-term in the chiral quark-soliton model}
\author{P.~Schweitzer$^{a}$, S.~Boffi$^{a,b}$, M.~Radici$^{a,b}$ \\
\footnotesize\it $^a$ 	Dipartimento di Fisica Nucleare e Teorica, 
			Universit\`a degli Studi di Pavia, I-27100 Pavia, 
			Italy\\
\footnotesize\it $^b$ 	Istituto Nazionale di Fisica Nucleare, 
			Sezione di Pavia, I-27100 Pavia, Italy}
\date{September 30, 2002}
\maketitle

\begin{abstract}
	Mellin moments of off-forward distribution functions are 
	even polynomials of the skewedness parameter $\xi$.
	This constraint, called polynomiality property,
	follows from  Lorentz- and time-reversal invariance.
	We prove that the unpolarized off-forward distribution functions in
	the chiral quark-soliton model satisfy the polynomiality property.
	The proof is an important contribution to the demonstration that
	the description of off-forward distribution functions in the model
	is consistent.
	As a byproduct of the proof we derive explicit model expressions 
	for moments of the $D$-term and compute the first coefficient in 
	the Gegenbauer expansion for this term.\\

\end{abstract}
%
\section{Introduction}\label{sect-1-introduction}

Off-forward distribution functions (OFDFs) of partons in the nucleon 
-- or equivalently double, skewed or non-diagonal parton distribution 
functions \cite{Muller:1998fv,Radyushkin:1996ru,Ji:1996ek,Collins:1996fb}
-- are a promising source of new information on the internal nucleon structure
(see Refs.~\cite{Ji:1998pc,Radyushkin:2000uy,Goeke:2001tz} for recent reviews).
OFDFs enter the description of hard exclusive processes, such as
hard meson production \cite{Collins:1996fb} and 
deeply virtual Compton scattering \cite{factorization-DVCS}.
Only recently first data on the latter process became available 
\cite{Airapetian:2001yk,Stepanyan:2001sm,Saull:1999kt,Favart:2001yj},
and so our present understanding of non-perturbative aspects of 
OFDFs has to rely on models.

After the first estimates in the framework of the MIT bag model 
\cite{Ji:1997gm}, studies in the chiral quark-soliton model ($\chi$QSM) 
of the nucleon have been presented \cite{Petrov:1998kf,Penttinen:1999th}.
The $\chi$QSM has been derived from the instanton model of the
QCD vacuum and provides a relativistic quantum field-theoretical
description of the nucleon. Without any adjustable parameters it describes
a large variety of nucleonic properties, like form factors 
\cite{Christov:1995hr,Christov:1995vm} and (anti)quark distribution 
functions \cite{Diakonov:1996sr,Diakonov:1997vc,Pobylitsa:1998tk} typically 
within $(10-30)\%$.

However, the reliability of a model is not only based on its
phenomenological success. From a theoretical point of view, it 
is important to demonstrate the internal consistency of the model.
One of the virtues of the $\chi$QSM, which is due to its field-theoretical 
character, is the possibility to provide analytical proofs which explicitly
demonstrate the consistency of the model.
In Ref.~\cite{Diakonov:1996sr} it has been proven that the quark and 
antiquark distribution functions computed in the model satisfy all 
general requirements, such as sum rules, positivity and inequalities.
With the same rigour it has been shown in 
Refs.~\cite{Petrov:1998kf,Penttinen:1999th} that the $\chi$QSM
expressions for OFDFs reduce to usual parton distributions 
in the forward limit, and that their first moments yield form factors.

In this paper we will present a further check of consistency for the $\chi$QSM,
and give for the unpolarized OFDFs an explicit proof of polynomiality, i.e. 
of the property that the $m^{\rm th}$ Mellin moment of an OFDF is an even 
polynomial in the skewedness variable $\xi$ of degree less than or equal to 
$m$ at the (unphysical) value $t=0$ of the Mandelstam variable~\footnote{Here
	we use the notation of Ref.~\cite{Ji:1996ek}.}. 
The polynomiality property follows from Lorentz- and
time-reversal invariance of strong interactions. 
The proof of polynomiality is the main result of this paper.

An elegant way to satisfy the polynomiality condition is to use the double 
distributions (DDs) of Refs.~\cite{Muller:1998fv,Radyushkin:1996ru}.
However, if assumed to be regular functions, DDs yield a vanishing highest 
power of $\xi$ for singlet moments of unpolarized OFDFs. 
This means that DDs yield an incomplete description, which can be completed 
by introducing the so-called $D$-term \cite{Polyakov:1999gs}.
Moments of the $D$-term are uniquely defined in terms 
of the coefficients of the highest power of $\xi$ 
\cite{Polyakov:1999gs,Belitsky:2000vk,Teryaev:2001qm}.
In Ref.~\cite{Kivel:2000fg} the $D$-term has been calculated numerically from 
moments of the OFDFs evaluated in the model for physical values of $t$.
The results obtained here enable us to evaluate in the model, in principle, 
all Mellin moments of the $D$-term directly at the unphysical point $t=0$.
We will demonstrate this for the lowest non-vanishing moment.

The paper is organized as follows. In 
Sec.~\ref{sect-2-model} 
	a brief introduction to the $\chi$QSM is given. In 
Sec.~\ref{sect-3-OFDF-in-model} the model expressions for the unpolarized 
	OFDFs are presented. The proof of polynomiality is given in
Secs.~\ref{sect-4-proof-H} and 
      \ref{sect-5-proof-E} for the non-spin-flip and the spin-flip OFDF.
Sec.~\ref{sect-6-D-term} is devoted to the $D$-term. Finally, in 
Sec.~\ref{sect-7-conclusions} we summarize our results and conclude.
Some technical details required for the proofs are given in the Appendices.

\section{The chiral quark-soliton model \boldmath ($\chi$QSM)}
\label{sect-2-model}

The $\chi$QSM \cite{Diakonov:1987ty} is essentially based on the principles 
of chiral symmetry breaking and the limit of a large number of colours $N_c$.
The effective chiral relativistic quantum field theory, which underlies the 
$\chi$QSM, is formulated in terms of quark, antiquark ($\psi$, $\bar\psi$) 
and Goldstone boson (pion $\pi^a$, $a=1,\,2,\,3$) degrees of freedom. 
It is given by the partition function 
\cite{Diakonov:tw,Dhar:gh,Diakonov:1985eg}
\be{
	Z_{\rm eff} = \int{\cal D}\psi\,{\cal D}\bar{\psi}\,{\cal D}U\;
	\exp\Biggl(i\int\di^4x\;\bar{\psi}\,
	(i\fslash{\partial}-M\,U^{\gamma_5})\psi\Biggr) \;\;,\;\;\;\; 
	U^{\gamma_5} = e^{i\gamma_5\tau^a\pi^a} .\label{eff-theory} }\ee
In Eq.~(\ref{eff-theory}) $M$ is the dynamical quark mass due to spontaneous 
breakdown of chiral symmetry, which is in general momentum dependent, and
$U=\exp(i\tau^a\pi^a)$ denotes the $SU(2)$ chiral pion field. 
The effective theory in Eq.~(\ref{eff-theory}) contains the Wess-Zumino term  
and the four-derivative Gasser-Leutwyler terms with correct coefficients.
It has been derived from the instanton model of the QCD vacuum
\cite{Diakonov:1985eg,Diakonov:1983hh}, which provides a 
mechanism of chiral symmetry breaking. 
The effective theory (\ref{eff-theory}) is valid at low energies below 
$600\,{\rm MeV}$, a scale set by the inverse of the average instanton size.

One application of the effective theory (\ref{eff-theory}) is the chiral 
quark-soliton model ($\chi$QSM) of baryons \cite{Diakonov:1987ty}.
In large-$N_c$ limit the path integral over pion field configurations in 
Eq.~(\ref{eff-theory}) can be solved in the saddle-point approximation.
The large-$N_c$ limit is known to be a good theoretical guideline.
In this limit the nucleon can be viewed as a classical soliton 
of the pion field \cite{Witten:tx}.
The $\chi$QSM provides a realization of this idea.
In the leading order of the large-$N_c$ limit the pion field is static,
and one can determine the spectrum of the effective one-particle Hamiltonian 
of the theory (\ref{eff-theory})
\be
	\hat{H}_{\rm eff}|n\ra = E_n |n\ra \;\;,\;\;\;\; 
	\hat{H}_{\rm eff} = -i\gamma^0\gamma^k\partial_k+\gamma^0MU^{\gamma_5}
	\;.\label{eff-Hamiltonian}\ee
The spectrum consists of an upper and a lower Dirac continuum, which are
distorted by the pion field as compared to continua of the free 
Dirac-Hamiltonian
\be
	\hat{H}_0|n_0\ra = E_{n_0}|n_0\ra \;\;,\;\;\;\; 
	\hat{H}_0 = -i\gamma^0\gamma^k\partial_k+\gamma^0 M
	\;,\label{free-Hamiltonian}\ee
and of a discrete bound state level of energy $E_{\rm lev}$, for a 
strong enough pion field of unity winding number.
By occupying the discrete level and the states of lower continuum each 
by $N_c$ quarks in an anti-symmetric colour state, one obtains a state 
with unity baryon number. 
The soliton energy $E_{\rm sol}$ is a functional of the pion field,
\be
	E_{\rm sol}[U] = N_c 
	\biggl(E_{\rm lev}+\singlesum{E_n<0}(E_n-E_{n_0})\biggr) 
	\;.\label{soliton-energy}\ee 
Minimization of $E_{\rm sol}[U]$ determines the self consistent solitonic 
pion field $U_c$. 
This procedure is performed for symmetry reasons in the so-called hedgehog 
ansatz $\pi^a({\bf x}) = \frac{x^a\!}{|{\bf x}|}\;P(|{\bf x}|)$ with the
radial (soliton profile) function $P(|{\bf x}|)$.
The nucleon mass $\Mn$ is given by $E_{\rm sol}[U_c]$.
Quantum numbers of the baryon like momentum, spin and isospin are 
described by considering zero modes of the soliton.
Corrections in $1/N_c$ can be included by considering time dependent 
pion field configurations.
The results of the $\chi$QSM respect all general counting rules of the 
large-$N_c$ phenomenology.

For the following it is important to note that the effective Hamiltonian 
$\hat{H}_{\rm eff}$ commutes with the parity operator $\hat{\Pi}$ and the 
grand-spin operator $\hat{\bf K}$, defined as the sum of quark angular 
momentum and isospin.
Thus $\hat{H}_{\rm eff}$, $\hat{\Pi}$, $\hat{\bf K}^2$ and ${\hat K}^3$
form a maximal set of commuting operators, and the single quark states 
$|n\ra$ in Eq.~(\ref{eff-Hamiltonian}) are characterized by the
quantum numbers parity $\pi$, $K$ and $M$ 
\be
	|n\ra = |E_n, \, \pi, \,K, \,M \ra 
	\;.\label{states} \ee
The $\chi$QSM allows to evaluate in a parameter-free way nucleon matrix 
elements of QCD quark bilinear operators as 
\ba
&&	\mbox{\hspace{-0.5cm}}
	\la N',{\bf P'}|\bar{\psi}(z_1)\Gamma\psi(z_2)|N,{\bf P }\ra\nonumber\\
&&	= A^{\mbox{\tiny$\Gamma$}}_{\mbox{\tiny$NN'$}}\;
	2\Mn N_c \singlesum{n, \rm occ}
	\int\!\!\di^3{\bf X}\:e^{i({\bf P'-P}){\bf X}}\,
	\bar{\Phi}_n({\bf z}_1-{\bf X})\Gamma\Phi_n({\bf z}_2-{\bf X})\,
	e^{iE_n(z^0_1-z^0_2)} + \dots 
	\label{matrix-elements} \ea
where for $z_1\neq z_2$ the insertion of the gauge link is understood on the 
LHS. The dots in Eq.~(\ref{matrix-elements}) denote terms subleading in the 
$1/N_c$ expansion, which we will not need here. 
In Eq.~(\ref{matrix-elements}) $\Gamma$ is some Dirac and flavour matrix, 
$A^{\mbox{\tiny$\Gamma$}}_{\mbox{\tiny$NN'$}}$ a constant depending on 
$\Gamma$ and the spin and flavour quantum numbers of the nucleon state 
$|N\ra=|S_3,T_3\ra$, and $\Phi_n({\bf x}) = \la{\bf x}|n\ra$ are the 
coordinate-space wave-functions of the single quark states $|n\ra$ defined 
in Eqs.~(\ref{eff-Hamiltonian},~\ref{states}).
The sum in Eq.~(\ref{matrix-elements}) goes over occupied levels $n$  
(i.e. $n$ with $E_n\le E_{\rm lev}$), and vacuum subtraction is 
implied for $E_n < E_{\rm lev}$ as in Eq.~(\ref{soliton-energy}).

Many static nucleonic observables -- like magnetic moments, electric 
polarizabilities, axial properties, etc. -- have been computed in the 
$\chi$QSM in the way sketched in Eq.~(\ref{matrix-elements}). 
The results were found in good agreement with data
(see Ref.~\cite{Christov:1995vm} for a review).
In particular the model describes data on electromagnetic form factors 
up to $|t| \sim 1\,{\rm GeV}^2$ within (10-30)$\%$ 
\cite{Christov:1995hr,Christov:1995vm}.
In Ref.~\cite{Diakonov:1996sr} it has been demonstrated that the model
can be applied to the description of twist-2 quark and anti-quark 
distribution functions of the nucleon.
The consistency of the approach has been shown by giving proofs that 
the model expressions satisfy all general requirements of QCD
\cite{Diakonov:1996sr}.
The distribution functions computed in the $\chi$QSM 
\cite{Diakonov:1996sr,Diakonov:1997vc,Pobylitsa:1998tk} refer to a low 
normalization scale of around $600\,{\rm MeV}$, and agree with 
parameterizations performed at comparably low scales \cite{Gluck:1994uf} 
within (10-30)$\%$.
 
The success of the $\chi$QSM in the parameter-free description of 
-- among others -- form factors and distribution functions encourages 
confidence into the predictions for OFDFs made in 
Refs.~\cite{Petrov:1998kf,Penttinen:1999th}.

\section{Off-forward distribution functions in the \boldmath $\chi$QSM}
\label{sect-3-OFDF-in-model}

The unpolarized quark off-forward distribution functions are defined as
\cite{Ji:1996ek}
\ba
&&	\mbox{\hspace{-1cm}}
	\int\!\frac{\di\lambda}{2\pi}\,e^{i\lambda x}\la {\bf P'},S_3'|
	\bar{\psi}_q(-\lambda n/2)\,\fslash{n}\,
	\psi_q(\lambda n/2)|{\bf P},S_3\ra
	\nonumber\\
&&	= H^q(x,\xi,t)\;\bar{U}({\bf P'},S_3')\fslash{n}U({\bf P},S_3)
 	+ E^q(x,\xi,t)\;\bar{U}({\bf P'},S_3')\,
	  \frac{i\sigma^{\mu\nu}n_\mu\Delta_\nu\!}{2\Mn}\,U({\bf P},S_3) 
	+ \dots  \label{def-1} \ea
where for brevity the gauge link is omitted and the scale dependence 
not indicated. The dots denote higher-twist contributions. 
The light-like vector $n^\mu$, the four-momentum transfer $\Delta^\mu$, 
the skewedness parameter $\xi$ and the Mandelstam variable $t$ are defined as
\be
  	n^2 = 0  		\; ,\;\;\;	
  	n(P'+P) = 2		\; ,\;\;\;
  	\Delta^\mu = (P'-P)^\mu	\; ,\;\;\;
  	n\Delta = -2\xi		\; ,\;\;\;
	t = \Delta^2		\; .\label{def-2} \ee
In Eq.~(\ref{def-1}) $x\in[-1,1]$ with the understanding that for negative 
$x$ Eq.~(\ref{def-1}) describes minus the OFDF of the antiquark.
When evaluating the above expressions in the $\chi$QSM in the large-$N_c$
limit, one has to note that $\Mn={\cal O}(N_c)$ is much larger than
spatial components of the momentum transfer $|\Delta^i|={\cal O}(N_c^0)$
which in turn is much larger than the zero-component of the momentum transfer 
$|\Delta^0|={\cal O}(N_c^{-1})$. 
The variables $x$ and $\xi$ are of ${\cal O}(N_c^{-1})$.
Choosing the 3-axis for the light-cone space direction, 
we have in the ``large-$N_c$ kinematics'' 
\be
	n^\mu = (1,0,0,-1)/\Mn 		\;,\;\;\;
	t     = -\bDelta^2 		\;,\;\;\;
	\xi   = -\Delta^3/(2\Mn) 	\;.\label{def-large-Nc-kinematics}\ee
Different flavour combinations of the OFDFs exhibit different behaviour
in the large-$N_c$ limit \cite{Goeke:2001tz}
\ba
	(H^u+H^d)(x,\xi,t) = N_c^2 f(N_cx,N_c\xi,t) \;, &&
	(E^u-E^d)(x,\xi,t) = N_c^3 f(N_cx,N_c\xi,t) \;, 
	\label{H-E-largeNc-large}\\
	(H^u-H^d)(x,\xi,t) = N_c\, f(N_cx,N_c\xi,t) \;, &&
	(E^u+E^d)(x,\xi,t) = N_c^2 f(N_cx,N_c\xi,t) \;. 
	\label{H-E-largeNc-small} \ea
The functions $f(u,v,t)$ in 
Eqs.~(\ref{H-E-largeNc-large},~\ref{H-E-largeNc-small}) are stable in the 
large-$N_c$ limit for fixed values of the  ${\cal O}(N_c^0)$ variables 
$u=N_cx$, $v=N_c\xi$ and $t$, and of course different for the different OFDFs.

The model expressions for the leading OFDFs, Eq.~(\ref{H-E-largeNc-large}),
have been derived in Ref.~\cite{Petrov:1998kf} and read
\ba
	(H^u+H^d)(x,\xi,t)&=& \Mn N_c \int\!\!\di^3{\bf X}\,e^{i\bDelta{\bf X}}
	\singlesum{n,\rm occ} 
	\int\!\frac{\di z^0}{2\pi}\, e^{iz^0(x\Mn - E_n)}\nonumber\\
	&&\times\phantom{\biggl|}
	\Phi^{\!\ast}_n({\bf X}+{\textstyle\frac{z^0}{2}}{\bf e^3})\,
	(1+\gamma^0\gamma^3)\,
	\Phi_n({\bf X}-{\textstyle\frac{z^0}{2}}{\bf e^3})
	\;,\label{def-Hu+Hd-model}\\
	&&\phantom{a}\nonumber\\
	(E^u-E^d)(x,\xi,t)&=& \,
	\frac{2i\Mn^2 N_c}{3(\bDelta^{\!\perp})^2}
	\int\!\!\di^3{\bf X}\,e^{i\bDelta{\bf X}}
	\singlesum{n,\rm occ} 
	\int\!\frac{\di z^0}{2\pi}\, e^{iz^0(x\Mn - E_n)} \nonumber\\
	&&\times\phantom{\biggl|}
	\Phi^{\!\ast}_n({\bf X}+{\textstyle\frac{z^0}{2}}{\bf e^3})\,
	(1+\gamma^0\gamma^3)(\btau\times\bDelta)^3
	\Phi_n({\bf X}-{\textstyle\frac{z^0}{2}}{\bf e^3})
	\;.\label{def-Eu-Ed-model} \ea
Here ${\bf e^3}$ is the unit vector of the 3-axis, which is singled 
out by the choices made in Eq.~(\ref{def-large-Nc-kinematics}), 
$\bDelta^{\!\perp}$ denotes the part of $\bDelta$ transverse to 3-axis,
and $(\btau\times\bDelta)^3\equiv \epsilon^{3kl}\tau^k\Delta^l$.

In Ref.~\cite{Petrov:1998kf} it has been demonstrated that in the forward limit
$(H^u+H^d)(x,\xi,t)$, Eq.~(\ref{def-Hu+Hd-model}), reduces to the 
model expression for unpolarized isoscalar distribution function
\be\label{forw-lim-Hu+Hd}
	\doublelim{\xi\to 0}{t\to0} (H^u+H^d)(x,\xi,t) = (f_1^u+f_1^d)(x)
\ee
and that the model expressions (\ref{def-Hu+Hd-model},~\ref{def-Eu-Ed-model}) 
are correctly normalized to the corresponding electromagnetic form factors
\be\label{norm-form-factors}
	\int\limits_{-1}^1\!\!\di x \;(H^u+H^d)(x,\xi,t) = (F_1^u+F_1^d)(t)
	\;\;,\;\;\;\;
	\int\limits_{-1}^1\!\!\di x \;(E^u-E^d)(x,\xi,t) = (F_2^u-F_2^d)(t)
	\;\;.\ee
The next and maybe most stringent check of the model expressions is the 
demonstration that Eqs.~(\ref{def-Hu+Hd-model},~\ref{def-Eu-Ed-model}) 
fulfil the polynomiality condition. As a consequence of Lorentz- and
time-reversal invariance the $m^{\rm th}$ Mellin moment $M^{(m)}(\xi,t)$ of 
an OFDF is an even polynomial in $\xi$ of degree less than or equal to $m$ 
at $t=0$
\ba
	M^{q\,(m)}_H(\xi,0) \equiv
	\int\limits_{-1}^1\!\!\di x\:x^{m-1}\,H^q(x,\xi,0)
	= h^{q\,(m)}_0 + h^{q\,(m)}_2 \xi^2 + \dots + 
	    \cases{h^{q\,(m)}_m    \xi^m    \!\!\!\!\! & for $m$ even\cr
		   h^{q\,(m)}_{m-1}\xi^{m-1}\!\!\!\!\! & for $m$ odd,} &&
	\label{def-polynom-Hu+Hd}\\
	M^{q\,(m)}_E(\xi,0) \equiv
	\int\limits_{-1}^1\!\!\di x\:x^{m-1}\,E^q(x,\xi,0)
	= e^{q\,(m)}_0 + e^{q\,(m)}_2 \xi^2 + \dots + 
	    \cases{e^{q\,(m)}_m    \xi^m    \!\!\! & for $m$ even\cr
		   e^{q\,(m)}_{m-1}\xi^{m-1}\!\!\! & for $m$ odd.}&&
	\label{def-polynom-Eu-Ed} \ea
Due to the spin $\frac12$ of the nucleon the highest coefficients of even 
moments of $H^q(x,\xi,t)$ and $E^q(x,\xi,t)$ are related to each other by
\be
	h^{q\,(m)}_m  =  - \,e^{q\,(m)}_m 
	\;\;.\label{relation-h-e} \ee
The point $t=0$ is unphysical and can be reached only by means of analytical 
continuation. 
In Ref.~\cite{Petrov:1998kf} $(H^u+H^d)(x,\xi,t)$ and $(E^u-E^d)(x,\xi,t)$
have been computed as functions of (physical values of) $x$, $\xi$ and $t$, 
and the polynomiality conditions, 
Eqs.~(\ref{def-polynom-Hu+Hd},~\ref{def-polynom-Eu-Ed}), have been checked
by taking (numerically) moments $M^{(m)}_H(\xi,t)$, $M^{(m)}_E(\xi,t)$
and extrapolating (numerically) to the unphysical point $t=0$
\cite{Kivel:2000fg}.
It is clear in this way one cannot prove strictly polynomiality.
This will be done in the next two sections.

\section{Proof of polynomiality for \boldmath $(H^u+H^d)(x,\xi,t)$}
\label{sect-4-proof-H}

The $m^{\rm th}$ moment of $(H^u+H^d)(x,\xi,t)$ in Eq.~(\ref{def-Hu+Hd-model})
-- which we will refer to as $M^{(m)}_H(\xi,t)$ -- reads
(see App.~\ref{App:moments})
\be
	M^{(m)}_H(\xi,t)
	=
	\frac{N_c}{\Mn^{m-1}} \singlesum{n,\rm occ}
	\sum\limits_{k=0}^{m-1}\binomial{m-1}{k}
	\frac{E_n^{m-1-k}\!\!\!}{2^k}\;
	\sum\limits_{j=0}^k \binomial{k}{j}
	 \la n|(1+\gamma^0\gamma^3)\,(\hat{p}^3)^{j}
	\exp(i\bDelta\hat{\bf X})\, (\hat{p}^3)^{k-j}|n\ra 
	\;,\label{Hu+Hd-model-mom1}\ee
where $\hat{p}^3$, $\hat{\bf X}$ mean the $3^{\rm rd}$ component of the 
free-momentum operator and the position operator, respectively.
The next step is to take the limit $t\to 0$ with $\xi\neq 0$ in 
$M^{(m)}_H(\xi,t)$. 
Eq.~(\ref{Hu+Hd-model-mom1}) shows that $M^{(m)}_H(\xi,t)=M^{(m)}_H(\bDelta)$,
i.e. $\bDelta$ determines entirely the dependence of the moments on 
$\xi$ and $t$ according to Eq.~(\ref{def-large-Nc-kinematics}).
Moreover, $\bDelta$ appears only in the operator $\exp(i\bDelta\hat{\bf X})$.
The order of the operations of taking the limit $t\to 0$ and taking matrix 
elements in Eq.~(\ref{Hu+Hd-model-mom1}) can be interchanged. 
Taking the limit $t\to 0$ (understood as analytical continuation) on
$\exp(i\bDelta\hat{\bf X})$ we obtain (see App.~\ref{App:an-cont-II})
\be
	\triplelim{\rm analytical}{\rm continuation}{t\to 0}
	\exp(i\bDelta\hat{\bf X}) = \sum\limits_{l_e=0}^\infty\,
	\frac{(-2i\xi\Mn|\hat{\bf X}|\,)^{l_e}}{l_e!\,}\; 
	P_{l_e}(\cos\hat{\theta}) 
	\;.\label{Hu+Hd-an-cont} \ee
Inserting Eq.~(\ref{Hu+Hd-an-cont}) into Eq.~(\ref{Hu+Hd-model-mom1}) 
and simplifying the result by using symmetries of the model 
(see App.~\ref{App:symmetries}) we obtain 
\ba
	M^{(m)}_H(\xi,0)
	&=&
	\frac{N_c}{\Mn^{m-1}} \singlesum{n,\rm occ}
	\sum\limits_{k=0}^{m-1}\binomial{m-1}{k}
	\frac{E_n^{m-1-k}\!\!\!}{2^k}\;
	\doublesumUp{l_e=0}{l_e\,\rm even}{\infty}\,
	\frac{(-2i\xi\Mn)^{l_e}}{l_e!\,} \nonumber\\
	&& \times \sum\limits_{j=0}^k\binomial{k}{j}
	\la n|(\gamma^0\gamma^3)^k\,(\hat{p}^3)^{j}
	|\hat{\bf X}|^{l_e}\,P_{l_e}(\cos\hat{\theta})
	\, (\hat{p}^3)^{k-j}|n\ra 
	\;,\label{Hu+Hd-model-mom2} \ea
where $(\gamma^0\gamma^3)^k$, equal to $\gamma^0\gamma^3$ for odd and unity
for even $k$, is introduced to simplify the notation.
Note that in the expression for $M^{(m)}_H(\xi,0)$ only even powers
of $\xi$ appear. Odd powers of $\xi$ vanish due to parity transformation
(see App.~\ref{App:symmetries}). 
What remains to be shown is that this series in $\xi^2$ is a polynomial
of degree less than or equal to $m$.

The operators in the matrix elements in Eq.~(\ref{Hu+Hd-model-mom2}) transform 
as irreducible tensor operators $\hat{T}^L_{\!M}$ of rank $L$ and $M=0$ under
simultaneous rotations in space and isospin-space. 
More precisely $|\hat{\bf X}|$ (and any power of it) is rank zero,
$\gamma^0\gamma^3$ and $\hat{p}^3$ are rank 1,
$P_{l_e}(\cos\hat{\theta})$ is rank $l_e$
(see App.~\ref{App:irreducible-tensor-op}).
The product $[\hat{T}_{\!1}]^{L_1}_{M_1}[\hat{T}_{\!2}]^{L_2}_{M_2}$ 
of two irreducible tensor operators $[\hat{T}_{\!1}]^{L_1}_{M_1}$ and 
$[\hat{T}_{\!2}]^{L_2}_{M_2}$ can be decomposed into a sum of 
(certain new) irreducible tensor operators according to 
\be
	[\hat{T}_{\!1}]^{L_1}_{M_1}[\hat{T}_{\!2}]^{L_2}_{M_2}
	= \sum\limits_{L=|L_1-L_2|}^{L_1+L_2} \hat{T}^{L}_{M}
	\;\;\;\mbox{with}\;\;\;\hat{T}^{L}_{M}\equiv C(LM;\,L_1M_1,L_2M_2)
	\; [\hat{T}_{\!1}]^{L_1}_{M_1}\;[\hat{T}_{\!2}]^{L_2}_{M_2} 
	\;,\label{Hu+Hd-Wig-Eck-0} \ee
where the $C(LM;\,L_1M_1,L_2M_2)$ denote Clebsch-Gordan coefficients.
The product of three or more irreducible tensor operators 
can be decomposed in an analogous way. 

For even $k$ in Eq.~(\ref{Hu+Hd-model-mom2}), there is a product of $k+1$
irreducible tensor operators, namely the rank-1 operator $\hat{p}^3$, 
which appears $k$-times, and the rank-$l_e$ operator 
$|\hat{\bf X}|^{l_e}\,P_{l_e}(\cos\hat{\theta})$.
For odd $k$ there is the further operator $\gamma^0\gamma^3$,
i.e. there are altogether $k+2$ irreducible tensor operators.
Upon successive application of Eq.~(\ref{Hu+Hd-Wig-Eck-0}) these $k+1$ or 
$k+2$ irreducible tensor operators yield  a series of certain new irreducible 
tensor operators $[\hat{T}_{\!H}]^L_M$ with even ranks $L$ ranging from zero 
to
\be
	L_{\rm max} = \cases{k+l_e+1 & if $k$ odd \cr 
	                     k+l_e   & if $k$ even.} 
	\label{Hu+Hd-Wig-Eck-1a}\ee
Note that only even ranks appear since all involved operators
have the ``magnetic quantum number'' $M=0$. 
The operators in Eq.~(\ref{Hu+Hd-model-mom2}) can thus be decomposed as 
\be
	(\gamma^0\gamma^3)^k\,(\hat{p}^3)^{j}
	|\hat{\bf X}|^{l_e}\,P_{l_e}(\cos\hat{\theta}) \, (\hat{p}^3)^{k-j}
	= \sum\limits_{L=0}^{L_{\rm max}} a_{L}^{\phantom{X}} \, 
	  [\hat{T}_{\!H}]^{L}_{0} 
	\;\;.\label{Hu+Hd-Wig-Eck-1b} \ee
Writing out explicitly the dependence of the single quark states 
$|n\ra$ on all quantum numbers, Eq.~(\ref{states}), we have
\ba
	&& \mbox{\hspace{-2cm}}
	\singlesum{n, \rm occ}
	\la n|\,(\gamma^0\gamma^3)^k\,(\hat{p}^3)^{j}
	|\hat{\bf X}|^{l_e}\,P_{l_e}(\cos\hat{\theta})\,(\hat{p}^3)^{k-j}|n\ra
	\nonumber\\
	&=&
	\singlesum{n, \rm occ}\;\singlesum{K,M}\;
	\sum\limits_{L=0}^{L_{\rm max}} \; a_{L}^{\phantom{X}} \;
	\la E_n, \,\pi, \,K, \,M |\, [\hat{T}_{\!H}]^{L}_{0} 
	|E_n, \, \pi, \,K, \,M \ra 
	\;\;.\label{Hu+Hd-Wig-Eck-1c} \ea 
On the RHS of Eq.~(\ref{Hu+Hd-Wig-Eck-1c}) we recognize traces 
(sums over matrix elements diagonal in $K$, $M$) 
of irreducible tensor operators, and the latter vanish unless 
the operator has rank zero \cite{Fano-Racah}.
As a consequence, the sum over $L$ in Eq.~(\ref{Hu+Hd-Wig-Eck-1c})
receives a non-zero contribution from $[\hat{T}_{\!H}]^{0}_{0}$ only.

Consider the case when $k$ is odd. 
The situation for $L=0$ in which $l_e$ takes its largest possible 
value $l_e^{\rm max}$ happens when $\gamma^0\gamma^3$ and the 
(in whatever way ordered) $k$ $\hat{p}^3$-operators combine to 
an operator of rank $k+1$. Then $l_e^{\rm max}=k+1$.
If $k$ is even, $(\gamma^0\gamma^3)^k=1$ does not contribute and then
$l_e^{\rm max}=k$. So, for a given $k$, the infinite series in $l_e$
in Eq.~(\ref{Hu+Hd-model-mom2}) actually stops at an $l_e^{\rm max}$
which is a function of $k$
\be
	l_e^{\rm max}(k) = \cases{k+1 & for $k$ odd \cr k  & for $k$ even.}
	\label{Hu+Hd-Wig-Eck-3} \ee
Inserting the result (\ref{Hu+Hd-Wig-Eck-3}) into 
Eq.~(\ref{Hu+Hd-model-mom2}) yields
\ba
	M^{(m)}_H(\xi,0)
	&=&
	\frac{N_c}{\Mn^{m-1}} \singlesum{n,\rm occ}
	\sum\limits_{k=0}^{m-1}\binomial{m-1}{k}
	\frac{E_n^{m-1-k}}{2^k}
	\doublesumUp{l_e=0}{l_e\,\rm even}{l_e^{\rm max}(k)}\,
	\frac{(-2i\xi\Mn)^{l_e}}{l_e!\,} \nonumber\\
	&& \times \sum\limits_{j=0}^k\binomial{k}{j}
	\la n|(\gamma^0\gamma^3)^k\,(\hat{p}^3)^{j}
	|\hat{\bf X}|^{l_e}\,P_{l_e}(\cos\hat{\theta})
	\, (\hat{p}^3)^{k-j}|n\ra 
	\;.\label{Hu+Hd-model-mom3} \ea
We observe that $M^{(m)}_H(\xi,0)$ is a polynomial in 
even powers of $\xi$, the highest power being given by
\be
	l_e^{\rm max}(k)\biggl|_{k=m-1} 
	= \cases{ m  & for $m$ even \cr  m-1 & for $m$ odd,}
	\label{Hu+Hd-Wig-Eck-4}\ee
which completes the proof of polynomiality, Eq.~(\ref{def-polynom-Hu+Hd}), 
for $(H^u+H^d)(x,\xi,t)$.

\section{Proof of polynomiality for \boldmath $(E^u-E^d)(x,\xi,t)$}
\label{sect-5-proof-E}

The model expression for the $m^{\rm th}$ moment of 
$(E^u-E^d)(x,\xi,t)$ in Eq.~(\ref{def-Eu-Ed-model})
is given by (see App.~\ref{App:moments})
\ba
	M^{(m)}_E(\xi,t)
	&=& \frac{2iN_c}{3\Mn^{m-2}} \singlesum{n,\rm occ}
	\sum\limits_{k=0}^{m-1}\binomial{m-1}{k}
	\frac{E_n^{m-1-k}\!\!\!}{2^k}\;
	\sum\limits_{j=0}^k\binomial{k}{j}\nonumber\\
	&&\times
	\la n|(1+\gamma^0\gamma^3)\,(\hat{p}^3)^{j}(\btau\times\bDelta)^3\,
	\frac{\exp(i\bDelta\hat{\bf X})}{(\bDelta^{\!\perp})^2}
	\, (\hat{p}^3)^{k-j}|n\ra 
	\;.\label{Eu-Ed-model-mom1}\ea
To perform the limit $t\to 0$ in $M^{(m)}_E(\xi,t)$ we have to continue 
analytically 
$i\bDelta^{\!\perp}\exp(i\bDelta\hat{\bf X})/(\bDelta^{\!\perp})^2$ to $t=0$.
The result reads (see App.~\ref{App:an-cont-II})
\be
	\triplelim{\rm analytical}{\rm continuation}{t\to 0,\,\xi\neq 0}
	\frac{i\bDelta^{\!\perp}
	\exp(i\bDelta\hat{\bf X})}{(\bDelta^{\!\perp})^2}
	= \sum\limits_{l_e = 1}^\infty\,\frac{(-2i\xi\Mn)^{l_e-2}}{l_e!\,}
	  \;i\biggl[\,\hat{\bf p}^{\perp}\;,\,
	  \;|\hat{\bf X}|^{l_e}P_{l_e}(\cos\hat{\theta}) \biggr] 
	\;.\label{Eu-Ed-an-cont}\ee
Inserting Eq.~(\ref{Eu-Ed-an-cont}) into Eq.~(\ref{Eu-Ed-model-mom1}) and 
simplifying the result by means of symmetries (see App.~\ref{App:symmetries})
we obtain 
\ba
	M^{(m)}_E(\xi,0)
	&=& \frac{2N_c}{3\Mn^{m-2}} \singlesum{n,\rm occ}
	\sum\limits_{k=0}^{m-1}\binomial{m-1}{k}
	\frac{E_n^{m-1-k}}{2^k}
	\doublesumUp{l_e=2}{l_e\;\,\rm even}{\infty}
	\frac{(-2i\xi\Mn)^{l_e-2}}{l_e!}\nonumber\\
	&&\times\sum\limits_{j=0}^k\binomial{k}{j}
	\la n|(\gamma^0\gamma^3)^{k+1}\,(\hat{p}^3)^{j}
	  \;i\biggl[\, (\btau\times\hat{\bf p})^3\; , \,
	  \;|\hat{\bf X}|^{l_e}P_{l_e}(\cos\hat{\theta}) \biggr]
	  \,(\hat{p}^3)^{k-j}|n\ra 
	\;.\label{Eu-Ed-model-mom2} \ea
Note that the contribution of $l_e=1$ in Eq.~(\ref{Eu-Ed-an-cont}) is singular 
at $\xi=0$, but it does not contribute in Eq.~(\ref{Eu-Ed-model-mom2}), such
that $M^{(m)}_E(\xi,0)$ is well defined for all values of $\xi$.
We observe that $M^{(m)}_E(\xi,0)$ is a series in even powers of $\xi$. 
Thus, what remains to be done again, is to demonstrate that this series is 
actually a polynomial of degree less than or equal to $m$ according to 
Eq.~(\ref{def-polynom-Eu-Ed}).

For this purpose we note that also $(\btau\times\hat{\bf p})^3$ 
is an irreducible tensor operator of rank 1 with $M=0$ 
(see App.~\ref{App:irreducible-tensor-op}), such that in 
Eq.~(\ref{Eu-Ed-model-mom2}) we deal with matrix elements 
of a product of irreducible tensor operators.
As argued in the previous section, these operators can be decomposed into a 
sum of certain new irreducible tensor operators $[\hat{T}_{\!E}]^L_M$ with 
$M=0$ and (even) ranks $L$ where $0\le L\le L_{\rm max}$ with
\be
	L_{\rm max}= \cases{k+l_e+1 & if $k$ odd \cr k+l_e+2 & if $k$ even.} 
	\label{Eu-Ed-Wig-Eck-1a} \ee
The operator in Eq.~(\ref{Eu-Ed-model-mom2}) contains a commutator of
$(\btau\times\hat{\bf p})^3$ and 
$|\hat{\bf X}|^{l_e}P_{l_e}(\cos\hat{\theta})$.
Focusing first on one ordering of these operators the decomposition 
into a sum of irreducible tensor operators, which we will call 
$[\hat{T}_{\!E}^{(1)}]^L_M$, looks as follows
\be
	(\gamma^0\gamma^3)^{k+1}\,(\hat{p}^3)^{j}\,
	(\btau\times\hat{\bf p})^3\; |\hat{\bf X}|^{l_e}\;
	P_{l_e}(\cos\hat{\theta})\;(\hat{p}^3)^{k-j}
	= \sum\limits_{L=0}^{L_{\rm max}} b_{L}^{(1)}\;
	  [\hat{T}_{\!E}^{(1)}]^{L}_{0} 
	\;\;, \label{Eu-Ed-Wig-Eck-1b} \ee
where the coefficients $b_{L}^{(1)}$ and the operators
$[\hat{T}_{\!E}^{(1)}]^{L}_{0}$ follow from the successive application 
of the tensor operator multiplication rule in Eq.~(\ref{Hu+Hd-Wig-Eck-0}).
However, only the $[\hat{T}_{\!E}]^{0}_{0}$ contributes to the sum 
over single quark states in Eq.~(\ref{Eu-Ed-model-mom2}).

There are many ways to construct a rank zero operator out of 
$(\gamma^0\gamma^3)^{k+1}$, $(\hat{p}^3)^{j}$, $(\btau\times\hat{\bf p})^3$,
$|\hat{\bf X}|^{l_e}\,P_{l_e}(\cos\hat{\theta})$ and $(\hat{p}^3)^{k-j}$.
The relevant one (for the proof) is the one which allows $l_e$ to 
take its maximal value $l_e^{\rm max}$. This case happens when the ranks of 
$(\gamma^0\gamma^3)^{k+1}$, $(\hat{p}^3)^{j}$, $(\btau\times\hat{\bf p})^3$ 
and $(\hat{p}^3)^{k-j}$ add up to $k+2$ for even $k$, or to $k+1$ for odd $k$.
Then $l_e^{\rm max}(k)$ takes the value
\be
	l_e^{\rm max}(k) = \cases{k+1 & for $k$ odd \cr  k+2 & for $k$ even.}
	\label{Eu-Ed-Wig-Eck-3}\ee
Considering the other ordering of $(\btau\times\hat{\bf p})^3$ 
and $|\hat{\bf X}|^{l_e}P_{l_e}(\cos\hat{\theta})$ in the commutator in
Eq.~(\ref{Eu-Ed-model-mom2}), one of course arrives at different coefficients 
$b_L^{(2)}$ and operators $[\hat{T}_{\!E}^{(2)}]^{L}_{0}$ in 
Eq.~(\ref{Eu-Ed-Wig-Eck-1b}), but at the same conclusion of
Eq.~(\ref{Eu-Ed-Wig-Eck-3}).
Applying the result (\ref{Eu-Ed-Wig-Eck-3}) to Eq.~(\ref{Eu-Ed-model-mom2}) 
we obtain
\ba
	M^{(m)}_E(\xi,0)
	&=& \frac{2N_c}{3\Mn^{m-2}} \singlesum{n,\rm occ}
	\sum\limits_{k=0}^{m-1}\binomial{m-1}{k}
	\frac{E_n^{m-1-k}}{2^k}
	\doublesumUp{l_e=2}{l_e\;\,\rm even}{l_e^{\rm max}(k)}
	\frac{(-2i\xi\Mn)^{l_e-2}}{l_e!}\nonumber\\
	&&\times\sum\limits_{j=0}^k\binomial{k}{j}
	\la n|(\gamma^0\gamma^3)^{k+1}\,(\hat{p}^3)^{j}
	  \;i\biggl[\, (\btau\times\hat{\bf p})^3\; , \,
	  \;|\hat{\bf X}|^{l_e}P_{l_e}(\cos\hat{\theta}) \biggr]
	  \,(\hat{p}^3)^{k-j}|n\ra 
	\;,\label{Eu-Ed-model-mom3} \ea
and observe that $M^{(m)}_E(\xi,0)$ is a polynomial in even 
powers of $\xi$ with the highest power given by
\be
	l_e^{\rm max}(k)-2\biggl|_{k=m-1} 
	= \cases{ m-1 & for $m$ odd \cr  m-2 & for $m$ even.}
	\label{Eu-Ed-Wig-Eck-4}\ee
The conclusion in Eq.~(\ref{Eu-Ed-Wig-Eck-4}) completes our proof of 
polynomiality, Eq.~(\ref{def-polynom-Eu-Ed}), for $(E^u-E^d)(x,\xi,t)$.
Noteworthy, for an even moment the highest power could be $m$
according to Eq.~(\ref{def-polynom-Eu-Ed}), but it is $m-2$ in the model. 
In the next section we will see that this is a consequence of general 
large-$N_c$ counting rules.

\section{\boldmath The $D$-term}
\label{sect-6-D-term}

The double distributions (DDs) of Refs.~\cite{Muller:1998fv,Radyushkin:1996ru}
have the advantage of satisfying the polynomiality condition automatically.
However, the description of non-forward nucleon matrix elements of two-body 
operators on the light cone in terms of DDs is incomplete,
if one assumes them to be regular functions~\footnote{In principle it 
	is possible to obtain a complete description (and to avoid the 
	introduction of the $D$-term) by using an alternative definition 
	of double distributions \cite{Belitsky:2000vk}.}.
This becomes manifest in the absence of the highest power $\xi^m$
in even (singlet) moments $m$ of $H^q(x,\xi,t)$ and $E^q(x,\xi,t)$. 
In other words, the coefficients $h^{q\,(m)}_m$ and $\,e^{q\,(m)}_m$
in Eq.~(\ref{relation-h-e}) vanish. 
The description can, however, be completed by introducing the $D$-term
$D^q(z)$, where $z=x/\xi$ \cite{Polyakov:1999gs}.
The $D$-term satisfies $D^q(z)=-D^q(-z)$, has a non-vanishing support for
$|z|<1$ only and is uniquely defined in terms of its moments 
\cite{Polyakov:1999gs,Belitsky:2000vk,Teryaev:2001qm}
\be
	\int_{-1}^1\!\di z\,z^{m-1}D^q(z) =
	\cases{ h^{q\,(m)}_m=-e^{q\,(m)}_m & for even $m$ \cr 0 & for odd $m$,}
	\label{def-D-term}\ee
with $h^{q\,(m)}_m$ and $\,e^{q\,(m)}_m$ as defined in 
Eqs.~(\ref{def-polynom-Hu+Hd},~\ref{def-polynom-Eu-Ed},~\ref{relation-h-e}).

The $D$-term exhibits the following large-$N_c$ behaviour \cite{Goeke:2001tz} 
\ba
&&	(D^u+D^d)(z) = N_c^2 f(z) \;,\nonumber\\
&&	(D^u-D^d)(z) = N_c\, f(z) \;,
	\label{D-term-Nc} \ea
where the functions $f(z)={\cal O}(N_c^0)$ are constants in the limit of 
large $N_c$ (and of course different for the different flavour combinations). 
It is convenient to expand the $D$-term in terms of Gegenbauer polynomials 
$C^\nu_k(z)$ for $\nu=\frac32$, the solutions of the leading order 
ERBL-evolution equations \cite{Efremov:1979qk,Lepage:1979zb}, which 
govern its dependence on the renormalization scale $\mu$,
\be
	D^q(z) = (1-z^2)\singlesumUp{k=1,3,5,\dots}{\infty} d^q_k C^{3/2}_k(z)
	\;.\label{D-term-Gegenbauer-series} \ee
The constant $d^q_1$ is of particular interest, as it determines the behaviour
of the $D$-term (and of OFDFs) in the asymptotic limit of renormalization 
scale $\mu\to\infty$ \cite{Goeke:2001tz}~\footnote{
	$D^q(z)$ mixes under evolution with the corresponding gluon $D$-term 
	$D^g(z)$. 
	Thus, in the flavour singlet case, it is strictly speaking 
	$\sum_q d_1^q + d_1^g$, which is scale independent and determines 
	the asymptotic behaviour of the $D$-term 
	\cite{Goeke:2001tz}.}.

In the literature the following phenomenological approach for modeling 
became popular. 
One models the regular part of DDs 
(relying on educated guesses, physical intuition) and 
supplements the description by adding the $D$-term 
(using often information from the $\chi$QSM \cite{Kivel:2000fg}). 
This approach yields DDs/OFDFs consistent with all general requirements
including polynomiality.
Many recent estimates and predictions for exclusive processes have been done 
in this way 
\cite{Kivel:2000fg,Vanderhaeghen:1999xj,Korotkov:2001zn,Belitsky:2001ns}.

The phenomenological interest motivates further studies in the model.
Let us discuss first the flavour-non-singlet combination $(D^u-D^d)(z)$.
From Eqs.~(\ref{Eu-Ed-model-mom3},~\ref{Eu-Ed-Wig-Eck-4}) we see 
\be
	\int\limits_{-1}^1\!\!\di z\;z^{m-1}(D^u-D^d)(z)=0 
	\;\;\;\forall \;m
	\;,\label{D-term-non-singlet} \ee
which means that also $(D^u-D^d)(z)=0$.
Note, however, that this is a zero at the order ${\cal O}(N_c^3)$.
We obtained this result from moments of $(E^u-E^d)(x,\xi,t)={\cal O}(N_c^3)$
according to Eq.~(\ref{H-E-largeNc-large}).
Due to the relations (\ref{relation-h-e},~\ref{def-D-term}) it is 
$(H^u-H^d)(x,\xi,t)={\cal O}(N_c)$ which dictates the large-$N_c$ behaviour 
of the flavour-non-singlet $D$-term in Eq.~(\ref{D-term-Nc}), 
and not $(E^u-E^d)(x,\xi,t)={\cal O}(N_c^3)$. 
Consequently, the $D$-term does not contribute to $(E^u-E^d)(x,\xi,t)$
at the order of $N_c$ considered here, which is just our observation 
in Eq.~(\ref{D-term-non-singlet}).
The result (\ref{D-term-non-singlet}) is thus a further
demonstration of the consistency of the model, which respects 
the general large-$N_c$ counting rules.

From Eq.~(\ref{Hu+Hd-model-mom3}) we read off the model expressions for 
moments of the flavour-singlet $(D^u+D^d)(z)$,
\ba
	\int\limits_{-1}^1\!\!\di z\;z^{m-1} (D^u+D^d)(z) 
	&=& 
	2 N_c\Mn \singlesum{n,\,\rm occ} 
	\la n|\,\gamma^0\gamma^3\,\hat{O}^{(m)}|n\ra \;,\nonumber\\
	\hat{O}^{(m)} 
	&\equiv& 
	\frac{i^m}{m!\,}\;\sum\limits_{j=0}^{m-1} \binomial{m-1}{j}
	(\hat{p}^3)^{j}|\hat{\bf X}|^{m}\,P_{m}(\cos\hat{\theta})
	\, (\hat{p}^3)^{m-j-1} \;,\;\;\mbox{$m$ even.} \label{D-term-moments} 
\ea
Recalling that $\Mn={\cal O}(N_c)$, we see that all moments of $(D^u+D^d)(z)$ 
are ${\cal O}(N_c^2)$ in the large-$N_c$ limit, in agreement
with the general counting rules in Eq.~(\ref{D-term-Nc}) also in this case.

In principle Eq.~(\ref{D-term-moments}) enables us to evaluate all 
moments of the flavour-singlet $D$-term. 
Here we shall restrict ourselves to the evaluation of the second, 
i.e. lowest non-vanishing, moment. 
It is related by $\int_{-1}^1\di z\,z \,D^q(z)=\frac{4}{5}\,d_1^q$ to the 
first coefficient $d_1^q$ in the Gegenbauer expansion in 
Eq.~(\ref{D-term-Gegenbauer-series}).
The expression for the  flavour singlet $d_1\equiv d_1^u+d_1^d$
follows from Eq.~(\ref{D-term-moments}) for $m=2$ and reads 
\be
	d_1 = -\,\frac54\,N_c\Mn \singlesum{n,\,\rm occ}\la n|\,
	\gamma^0\gamma^3\, \biggl\{\hat{p}^3\, ,\;|\hat{\bf X}|\,
	P_2(\cos\hat{\theta})\biggr\} |n\ra 
	\;.\label{D-term-moment-1.1} \ee
We evaluate $d_1$ by means of the so-called interpolation formula 
\cite{Diakonov:1996sr}, which consists in exactly evaluating the contribution 
to $d_1$ in Eq.~(\ref{D-term-moment-1.1}) from the discrete level 
$d_1^{\rm lev}$, and estimating the contribution from the negative Dirac 
continuum, $d_1^{\rm cont}$, as follows.
One rewrites $d_1^{\rm cont}$ in terms of the Feynman quark propagator in the
static background pion field $U$, expands it in a series of gradients of the 
$U$-field, and evaluates the leading contribution.
The interpolation formula yields exact results in three limiting cases: 
(i) low momenta, $|\nabla U|\ll M$, 
(ii) large momenta, $|\nabla U|\gg M$, 
(iii) any momenta but small pion field, $|\log U|\ll 1$.
One can thus expect that it yields reliable estimates also for the 
general case. 
Indeed, it has been observed that estimates obtained in this way agree 
with results from exact and much more involved numerical calculations
within $(10-20)\%$ \cite{Diakonov:1996sr,Diakonov:1997vc,Pobylitsa:1998tk}.

The expectation value of the operator in Eq.~(\ref{D-term-moment-1.1}) 
vanishes in the grand-spin $K=0$ and positive parity state, which 
characterizes the discrete level, i.e. we have
\be
	d_1^{\rm lev} = 0 
	\;.\label{D-term-moment-1.2} \ee
For the contribution of $d_1^{\rm cont}$ we obtain
\ba
	d_1^{\rm cont} = -\frac{5}{4}\;f_\pi^2\Mn
	\int\!\!\di^3{\bf x}\; P_2(\cos\theta)\,{\bf x}^2\;
	\trF[\nabla^3 U]\,[\nabla^3U^\dag] 
	\;,\label{D-term-moment-1.3}\ea 
where $\trF$ denotes the trace over flavour.
In the calculation leading to Eq.~(\ref{D-term-moment-1.3})
we took the (generally momentum dependent) mass $M$ constant for simplicity.
This simplification is legitimate for finite (or at most only logarithmically
divergent) quantities, which do not (or only weakly do) depend on
regularization.

For constant $M$ the continuum contribution $d_1^{\rm cont}$ is 
logarithmically divergent and has to be regularized.
The same type of divergence appears in the model expression for the 
pion decay constant $f_\pi$. Regularizing both quantities consistently in the 
same scheme allows to eliminate regulator and cutoff in favour of $f_\pi$. 
It should be noted that further terms, containing three or more gradients of 
the $U$-field and neglected in Eq.~(\ref{D-term-moment-1.3}), are 
finite~\footnote{
	For more technical details on the interpolation formula, 
	and examples of similar calculations see 
	Refs.~\cite{Diakonov:1996sr,Diakonov:1997vc}.}. 

In the hedgehog-ansatz the $U({\bf x})$-field is specified 
by the soliton profile $P(r)$ with $r=|{\bf x}|$.
We take for $P(r)$ the analytical arctan-profile
\be
	P(r) = -2{\rm arctan}\frac{R^2}{r^2}\;,\;\;\;
	R = M^{-1} = (350\,{\rm MeV})^{-1}
	\;,\label{arctan-profile}\ee
which is known to well approximate the self-consistent profile.
In Eq.~(\ref{arctan-profile}) $R$ is the so-called size of the soliton. 
This enables us to continue the calculation analytically, and yields
\be\label{D-term-d1}
	d_1^{\rm cont} = -\;3\,\pi^2\sqrt{2}\;f_\pi^2\Mn R^3 \;. \ee
With $f_\pi = 91\,{\rm MeV}$ and the value $\Mn = 1170\;{\rm MeV}$ obtained 
in the $\chi$QSM with the analytical arctan-profile (\ref{arctan-profile})
-- and ``adding up'' the zero of Eq.~(\ref{D-term-moment-1.2}) -- we obtain 
\be
	d_1 = - 9.46 \;\;\mbox{at $\mu^2 \simeq (600\;{\rm MeV})^2$,}
	\label{D-term-d1a}\ee
the low scale of the model.
In order to understand how this result can be used for phenomenology, we write
down the entire sum rule for the second moment of $(H^u+H^d)(x,\xi,t)$ which
reads
\be
	\int\limits_{-1}^1\!\!\di x\;x\,(H^u+H^d)(x,\xi,0) 
	= M_Q + \frac45\;d_1\,\xi^2 
	\;,\label{D-term-sum-rule-2mom} \ee
where $M_Q$ is the fraction of the nucleon momentum carried by 
quarks and antiquarks.
In the $\chi$QSM at its low scale $M_Q$ is exactly unity since there are no 
gluons in the model and the entire nucleon momentum is carried by quarks and 
antiquarks \cite{Diakonov:1996sr}.
Considering ratios of model quantities to be more reliable than
absolute numbers, the prediction of the $\chi$QSM is 
\be
	\frac{d_1}{M_Q} = - 9.46 
	\;\;\mbox{at $\mu^2 \simeq (600\;{\rm MeV})^2$.}
	\label{D-term-identify-ratios} \ee
Neglecting the scale dependence of the ratio $d_1/M_Q$
(and this we can do since the effects of evolution are smaller than the 
intrinsic accuracy of the model result in Eq.~(\ref{D-term-identify-ratios}))
we obtain with $M_Q\simeq 0.5$ at a few ${\rm GeV}^2$ the
prediction from the $\chi$QSM for experimentally relevant scales
\be
	d_1 \simeq -4.7 \;\;\mbox{at few ${\rm GeV}^2$.} 
	\label{D-term-d1-prediction} \ee
The result in Eq.~(\ref{D-term-d1-prediction}) agrees well with the number
$d_1\approx -4.0$ obtained from the numerical results on $(H^u+H^d)(x,\xi,t)$
of Ref.~\cite{Petrov:1998kf} and quoted in Ref.~\cite{Kivel:2000fg}~\footnote{
	By numerically taking moments of $(H^u+H^d)(x,\xi,t)$ from 
	Ref.~\cite{Petrov:1998kf} and extrapolating to $t=0$ the value 
	$d_1/M_Q \approx -8.0$ has been found at the low scale of the model
	\cite{Polyakov-private-communication}.
	Neglecting the effects of evolution and using the phenomenological 
        value $M_Q\approx 0.5$ in Ref.~\cite{Kivel:2000fg} it was estimated 
	$d_1\approx -4.0$ at experimentally relevant scales of few 
	${\rm GeV}^2$ in Ref.~\cite{Kivel:2000fg}.}.
The difference between the two numbers is due to the use of different 
regularizations. 
In Ref.~\cite{Petrov:1998kf} the momentum dependent mass $M(p)$ was used
as a regulator. Here we took constant $M$ 
(and applied some regularization scheme, the dependence on 
which has been then removed in favour of $f_\pi$). 

The advantage of our method is that we are in a position to exactly evaluate,
at least in principle, any moment of the $D$-term in the model.
In Ref.~\cite{Kivel:2000fg,Polyakov-private-communication} this was not 
possible, not even in principle, due to limitations set by numerical accuracy.

\section{Summary and conclusions}
\label{sect-7-conclusions}

We have presented explicit proofs that the $\chi$QSM expressions for
the leading flavour combinations in the large-$N_c$ limit of 
the unpolarized OFDFs $(H^u+H^d)(x,\xi,t)$ and $(E^u-E^d)(x,\xi,t)$
satisfy the polynomiality condition.
The method can straightforwardly be generalized to the case of flavour 
combinations subleading in $N_c$, and to the case of helicity dependent OFDFs.
The proof is an important contribution to the demonstration of the
theoretical consistency of the description of OFDFs in the framework
of the $\chi$QSM, and increases the credibility and reliability of
theoretical predictions made on the basis of the $\chi$QSM.

As a byproduct of the proof, we derived explicit model expressions
for the moments of the flavour-singlet $D$-term,
which are of phenomenological and theoretical interest. 
The explicit model expressions not only shed some light on the
chiral dynamics underlying the $D$-term, but also enable exact model 
calculations of the coefficients in the Gegenbauer expansion of the $D$-term.
We have demonstrated it by analytically computing the first coefficient $d_1$
for the flavour-singlet $D$-term reproducing the numerically estimated value 
previously obtained. 
Calculations of higher Gegenbauer coefficients are in progress.

The $G_5$-symmetry transformation used in App.~\ref{App:symmetries}
is the ``non-standard'' time reversal transformation \cite{Weinberg+Wigner}
for quark fields in a static background pion field. We became aware of 
that after this work was completed thanks to Ref.~\cite{Anselmino:2002yx}.
Thus, as in QCD, the proof of the polynomiality property of OFDFs in the 
chiral quark-soliton model also relies on the principles of time reversal 
invariance (and hermiticity) and Lorentz-invariance, with the latter reduced 
to rotational invariance, once the soliton rest frame has been chosen to 
evaluate the model expressions for OFDFs.
More precisely, due to the specific hedgehog ansatz for the soliton field, 
the model expressions are invariant under simultaneous rotations in space 
and flavour space. 
It is not surprizing that the deep connection between flavour, space-time and 
chiral symmetry becomes apparent most easily here, in a relativistic chiral 
model of the nucleon based on the large-$N_c$ limit.

\vspace{0.5cm}
{\footnotesize
We would like to thank M.~V.~Polyakov and M.~Schwamb for discussions.
This work has partly been performed under the contract  
HPRN-CT-2000-00130 of the European Commission.}

\appendix
\section{Mellin moments of \boldmath $(H^u+H^d)$ and $(E^u-E^d)$} 
\label{App:moments}

$H(x,\xi,t)\equiv(H^u+H^d)(x,\xi,t)$ and
$E(x,\xi,t)\equiv(E^u-E^d)(x,\xi,t)$,
Eqs.~(\ref{def-Hu+Hd-model},~\ref{def-Eu-Ed-model}), 
can be written compactly as
\be
	A(x,\xi,t) =  c_{A}^{\phantom{X}} \singlesum{n\,\rm occ}
	\int\!\frac{\di z^0}{2\pi}\,e^{iz^0(x\Mn-E_n)}\,
	f_n(\Gamma_{\!A},z^0,\bDelta) 
	\label{App:mom-0}\ee
with
\ba
	c_{H}^{\phantom{X}} = N_c\Mn \;,\mbox{\hspace{0.7cm}}\; 
&& 	c_{E}^{\phantom{X}} = (2/3) i N_c\Mn^2/
			      (\bDelta^{\!\perp})^2 \;, \nonumber\\
	\Gamma_{\!H}=(1+\gamma^0\gamma^3) \;, 
&&	\Gamma_{\!E}=(1+\gamma^0\gamma^3)(\btau\times\bDelta)^3 \;,
	\label{App:mom-0a}\ea
and $f_n(\Gamma,z^0,\bDelta)$ defined as
\ba
	f_n(\Gamma,z^0,\bDelta) 
	&\equiv& 
	\int\!\!\di^3{\bf X}\,e^{i\bDelta{\bf X}} 
	\Phi^{\!\ast}_n({\bf X}+\frac{z^0}{2}{\bf e^3})\,\Gamma
	\Phi_n({\bf X}-\frac{z^0}{2}{\bf e^3}) \nonumber\\
	&=& \la n|\,\Gamma\,\exp(-i\frac{z^0}{2}\hat{p}^3)\,
	\exp(i\bDelta\hat{\bf X})\, \exp(-i\frac{z^0}{2}\hat{p}^3)|n\ra\;, 
	\label{App:mom-1}\ea
where in the second line we made use of $\Phi_n({\bf X}-{\bf a}) = 
\la{\bf X}-{\bf a}|n\ra=\la{\bf X}|\,\exp(-i{\bf a}\hat{\bf p})|n\ra$
and analogously for $\Phi^\ast_n({\bf X}-{\bf a})$.
For the $m^{\rm th}$ moment $M^{(m)}_A(\xi,t)$ we obtain
\ba
	M^{(m)}_A(\xi,t)
	&\equiv& \int\limits_{-1}^1\!\!\di x\;x^{m-1}\;A(x,\xi,t)
	= c_{A}^{\phantom{X}} \singlesum{n\,\rm occ} 
	\int\limits_{-1}^1\!\!\di x\;x^{m-1}
	\int\!\frac{\di z^0}{2\pi}\,e^{iz^0(x\Mn-E_n)}\,
	f_n(\Gamma_{\!A},z^0,\bDelta) \nonumber\\
	&=&
	\frac{c_{A}^{\phantom{X}}}{\Mn^{m}}
	\singlesum{n\,\rm occ}\int\!\frac{\di z^0}{2\pi}\, 
	\Biggl[\frac{\partial^{m-1}}{\partial (iz^0)^{m-1}}
	\int\limits_{-\infty}^\infty\!\!\di y\;e^{iz^0y}\Biggr]\,
	e^{-iz^0E_n}\,f_n(\Gamma_{\!A},z^0,\bDelta) \nonumber\\
	&=& 
	\frac{c_{A}^{\phantom{X}}}{\Mn^{m}} \singlesum{n\,\rm occ} 
	\Biggl[\frac{\partial^{m-1}}{\partial(-iz^0)^{m-1}}\;
	e^{-iz^0E_n} f_n(\Gamma_{\!A},z^0,\bDelta)\Biggr]_{z^0=0}\nonumber\\
	&=& 
	\frac{c_{A}^{\phantom{X}}}{\Mn^{m}}\singlesum{n\,\rm occ} 
	\sum\limits_{k=0}^{m-1} \binomial{m-1}{k} E_n^{m-1-k}
	\Biggl[\frac{\partial^k f_n(\Gamma_{\!A},z^0,\bDelta)}
		    {\partial(-iz^0)^k }\Biggr]_{z^0=0}  \;.
	\label{App:mom-2a}\ea
In the second line of Eq.~(\ref{App:mom-2a}),
after the substitution $y=x\Mn$, the new integration limits $[-\Mn,\,\Mn]$
have been replaced by $[-\infty,\,\infty]$. This step is justified in the
large-$N_c$ limit where $\Mn={\cal O}(N_c)$. 
The differentiation of $f_n(\Gamma,z^0,\bDelta)$ in the 
last line of Eq.~(\ref{App:mom-2a}) yields
\be
	\Biggl[\frac{\partial^k f_n(\Gamma,z^0,\bDelta)}
	            {\partial(-iz^0)^k}\Biggr]_{z^0=0}
	= \frac{1}{2^k}\sum\limits_{j=0}^k\binomial{k}{j}
	\la n|\Gamma\,(\hat{p}^3)^j
	\exp(i\bDelta\hat{\bf X})\, (\hat{p}^3)^{k-j}|n\ra \;. 
	\label{App:mom-2b}\ee
Inserting Eq.~(\ref{App:mom-2b}) into Eq.~(\ref{App:mom-2a}) we obtain 
\be
	M^{(m)}_A(\xi,t) =
	\frac{c_{A}^{\phantom{X}}}{\Mn^m} \singlesum{n,\rm occ}
	\sum\limits_{k=0}^{m-1}\binomial{m-1}{k}
	\frac{E_n^{m-1-k}\!\!\!}{2^k}\;
	\sum\limits_{j=0}^k\binomial{k}{j}
	 \la n|\Gamma_{\!A}\,(\hat{p}^3)^{j}
	\exp(i\bDelta\hat{\bf X})\, (\hat{p}^3)^{k-j}|n\ra  
	\;,\label{App:mom-2c}\ee
which is the result quoted in Eq.~(\ref{Hu+Hd-model-mom1}) 
and in Eq.~(\ref{Eu-Ed-model-mom1}), respectively.

\section{Analytical continuation to \boldmath $t=0$}
\label{App:an-cont-II}

The moments $M_H(\xi,t)$ and $M_E(\xi,t)$ in Eq.~(\ref{App:mom-2c}) depend on 
the variables $\xi$ and $t$ only through the 3-momentum transfer $\bDelta$ 
in the ``large-$N_c$ kinematics'' of Eq.~(\ref{def-large-Nc-kinematics}).
As only $\xi$ and $t$ are relevant variables, $\bDelta$ can be expressed 
through $\xi$, $t$ and some arbitrary real angle $\alpha$ as
\be
	\bDelta = \biggl( 
	-\sin\alpha\sqrt{-t-(2\xi\Mn)^2\,}\, ,\;
	 \cos\alpha\sqrt{-t-(2\xi\Mn)^2\,}\, ,\; -\,(2\xi\Mn)\biggr)
	\;.\label{App:an-cont-II-1}\ee
The arbitrariness of $\alpha$ reflects the azimuthal symmetry of the physical
situation around the 3-axis chosen for the space direction of the light cone
in Eq.~(\ref{def-large-Nc-kinematics}).
Since there is no real dependence on this angle, we can choose a 
particular value for $\alpha$, or equivalently average over it.

\paragraph*{(i) $(H^u+H^d)$.}
To continue the moments $M_H(\xi,t)$ to $t=0$, we have to continue 
analytically the function $F_H(\xi,t)\equiv\exp(i\bDelta{\bf X})$ in
Eq.~(\ref{Hu+Hd-model-mom1}). 
Taking $\bDelta$ as defined in Eq.~(\ref{App:an-cont-II-1}) we obtain 
a function $F_H(\xi,t)$ depending in addition on the angle $\alpha$, 
and we remove the dependence on this arbitrary angle by averaging 
over $\alpha\in[0,2\pi]$. Taking 
${\bf X}=|{\bf X}|(\sin\theta\,\cos\phi,\,\sin\theta\,\sin\phi,\,\cos\theta)$
in spherical coordinates and expanding $\exp(i\bDelta{\bf X})$ in a series,
we obtain for $F_H(\xi,t)$
\be
	F_H(\xi,t) \equiv \exp(i\bDelta{\bf X}) = \sum\limits_{l=0}^\infty
	\frac{(-i|{\bf X}|)^l\!}{l!}\;
	\frac{1}{2\pi}\!\int\limits_0^{2\pi}\!\!\di\alpha\,
	\biggl(\sqrt{-t-(2\xi\Mn)^2}\,\sin\theta\sin(\alpha-\phi)
	+(2\xi\Mn)\,\cos\theta\biggr)^l
	\;.\label{App:an-cont-II-2} \ee
We see that $F_H(\xi,t)$ is an analytic function of the variables $\xi$ and 
$t$ everywhere except for the twofold branching point $t=-(2\xi\Mn)^2$.
Taking the limit $t\to 0$ within a definite Riemann sheet, we obtain
\be
	\triplelim{\rm analytical}{\rm continuation}{t\to 0}  
	F_H(\xi,t) = \sum\limits_{l=0}^\infty
	\frac{(-i 2\xi\Mn|{\bf X}|)^l\!}{l!}\;
	\frac{1}{2\pi}\!\int\limits_0^{2\pi}\!\!\di\alpha\,
	\biggl(\pm i\,\sin\theta\sin(\alpha-\phi)+\cos\theta\biggr)^l
	\label{App:an-cont-II-3} \ee
with the plus and minus signs referring to the two Riemann sheets.
The final result is real and does not depend on the choice of sheet
\be
	\frac{1}{2\pi}\int\limits_{0}^{2\pi}\!\!\di\alpha\;
	\biggl(\pm i\sin\theta\sin(\alpha-\phi)+\cos\theta\biggr)^l 	
	= \frac{1}{\pi}\,\int\limits_0^{\pi}\!\!\di\alpha\;
	  \biggl(\pm i\sin\theta\sin\alpha+\cos\theta\biggr)^l
	= P_l(\cos\theta) 
	\;.\label{App:an-cont-II-4} \ee
One arrives at the intermediate result in Eq.~(\ref{App:an-cont-II-4}) 
by using the periodicity of the sine and cosine functions.
The last equality is the
{\sl integral representation of Laplace and Mehler for Legendre polynomials}
\cite{Magnus+Oberhettinger}. Inserting Eq.~(\ref{App:an-cont-II-4}) 
in  Eq.~(\ref{App:an-cont-II-3}) we arrive at the result quoted in 
Eq.~(\ref{Hu+Hd-an-cont}),
\be
	\triplelim{\rm analytical}{\rm continuation}{t\to 0}  F_H(\xi,t) 
	= \sum\limits_{l=0}^\infty \frac{(-i 2\xi\Mn|{\bf X}|)^l\!}{l!}\;
	P_l(\cos\theta)
	\;.\label{App:an-cont-II-5} \ee

\paragraph*{(ii) $(E^u-E^d)$.} 
From Eq.~(\ref{Eu-Ed-model-mom1}) we see that the function $F_E(\xi,t)$ 
-- which is to be continued to the point $t=0$ in that case -- reads
\be
	F_E(\xi,t)\equiv 
	\frac{i\bDelta^{\!\perp}\exp(i\bDelta{\bf X})}{(\bDelta^{\!\perp})^2}
	=\bnabla^\perp\;\frac{F_H(\xi,t)}{-t-(2\xi\Mn)^2} 
	\label{App:an-cont-II-6} \ee
with $F_H(\xi,t)$ from Eq.~(\ref{App:an-cont-II-2}).
In Eq.~(\ref{App:an-cont-II-6}) $\bnabla^\perp\equiv$ 
$(\frac{\partial\;}{\partial X^1},\frac{\partial\;}{\partial X^2})$.
For $\xi\neq 0$ the limit $t\to 0$ yields
\be
	\triplelim{\rm analytical}{\rm continuation}{t\to 0,\;\xi\neq0}	
	F_E(\xi,t) = \bnabla^\perp\; \sum\limits_{l=1}^\infty\,
	\frac{\;(-i2\xi\Mn|{\bf X}|)^{l-2}}{l!} \;P_l(\cos\theta) 
	\;,\label{App:an-cont-II-7} 
\ee
where we took the limit independently on numerator and denominator, since 
both limits exist, and used the result (\ref{App:an-cont-II-5}).
The $l=0$ addendum in the sum in Eq.~(\ref{App:an-cont-II-7}) is a constant 
and vanishes upon differentiation, so the sum starts with $l=1$. 
In the next Appendix we will see that the addendum for $l=1$, 
which is singular at $\xi=0$, does not contribute. Thus the 
moments $M^{(m)}_E(\xi,0)$ are well defined for all $\xi$.

Using the identity $\frac{\partial\;}{\partial\hat{\bf X}}\,F(\hat{\bf X})=$
$i[\hat{\bf p},F(\hat{\bf X})]$ in Eq.~(\ref{App:an-cont-II-7}) we obtain the 
result quoted in Eq.~(\ref{Eu-Ed-an-cont}).

\section{Symmetries of the model} 
\label{App:symmetries}

\paragraph*{$G_5$-symmetry.}
Consider the unitary matrix $G_5$ with the following property
\be
 	G_5\gamma^\mu G_5^{-1} = (\gamma^\mu)^T  \;\;,\;\; 
	G_5\tau^a     G_5^{-1} = -(\tau^a)^T 	 \;\;.\label{App:sym-G5}
\ee
In the standard representation of $\gamma$- and $\tau$-matrices  
$G_5 = \gamma^1\gamma^3\tau^2$.
Noting that $\nabla^i=-(\nabla^i)^T$ holds formally, one finds that $G_5$
transforms the Hamiltonian $\hat{H}_{\rm eff}$ and the single quark states,
Eqs.~(\ref{eff-Hamiltonian},~\ref{states}), in coordinate-space representation 
as $G_5 \hat{H}_{\rm eff}  G_5^{-1} = \hat{H}_{\rm eff}^T$ and
$G_5 \Phi_n({\bf x}) = \Phi_n^\ast({\bf x})$ \cite{Christov:1995vm}.

Let $\Gamma$ be some matrix in Dirac- and flavour-space and define
$\tilde{\Gamma}\equiv(G_5\Gamma G_5^{-1})^T$. Let $F(\hat{\bf X})$
be a function of $\hat{\bf X}$. Then 
\be
	\la n|\Gamma(\hat{p}^3)^l F(\hat{\bf X}) (\hat{p}^3)^m|n\ra
	= (-1)^{l+m} \la n|\tilde{\Gamma}
	(\hat{p}^3)^m F(\hat{\bf X}) (\hat{p}^3)^l|n\ra\ph 
	\;.\label{App:sym-1}
\ee

\paragraph*{Parity.}
The parity operator $\hat{\Pi}=\hat{\Pi}^{-1}\equiv\gamma^0\hat{\cal P}$,
where $\hat{\cal P} F(\hat{\bf X}) \hat{\cal P}^{-1} =  F(-\hat{\bf X})$,
acts on the single quark states as $\hat{\Pi} |n\ra = \pm |n\ra$ 
according to the parity of the single quark state in Eq.~(\ref{states}). 
With the notation of Eq.~(\ref{App:sym-1}) upon application of the parity 
transformation we have
\be
 	\la n|\Gamma(\hat{p}^3)^l F(\hat{\bf X}) (\hat{p}^3)^m|n\ra
	= (-1)^{l+m} \la n|\left(\gamma^0\Gamma\gamma^0\right)(\hat{p}^3)^l
	  \left(\hat{\cal P}F(\hat{\bf X})\hat{\cal P}^{-1}\right)
	  (\hat{p}^3)^m|n\ra 
	\;.\label{App:sym-2}
\ee

\paragraph*{(i) $(H^u+H^d)$.} 
Here $\Gamma=(1+\gamma^0\gamma^3)$ and $\tilde{\Gamma}=(1-\gamma^0\gamma^3)$,
and $F(\hat{\bf X})=|\hat{\bf X}|^{l_e} \hat{P}_{l_e}$ where
$\hat{P}_l\equiv P_l(\cos\hat{\theta})$ is set for brevity.
According to Eq.~(\ref{App:sym-1})
\be
	\sum\limits_{j=0}^k\binomial{k}{j}
	\la n|(1+\gamma^0\gamma^3)(\hat{p}^3)^j
	|\hat{\bf X}|^l \hat{P}_l (\hat{p}^3)^{k-j}|n\ra
	= (-1)^k \sum\limits_{j=0}^k\binomial{k}{j}
	\la n|(1-\gamma^0\gamma^3)\,(\hat{p}^3)^{k-j}
	|\hat{\bf X}|^l \hat{P}_l\,(\hat{p}^3)^j|n\ra 
	\;.\label{App:sym-Hu+Hd-1}
\ee
From Eq.~(\ref{App:sym-Hu+Hd-1}) it follows that 
the contribution of the ``1'' in $(1+\gamma^0\gamma^3)$ vanishes for odd $k$,
while the contribution of ``$\gamma^0\gamma^3$'' vanishes for even $k$.
Using $(\gamma^0\gamma^3)^k=1$ for even $k$, and 
$(\gamma^0\gamma^3)^k=\gamma^0\gamma^3$ for odd $k$, 
this conclusion can be written compactly as
\be
	\sum\limits_{j=0}^k\binomial{k}{j}
	\la n|(1+\gamma^0\gamma^3)(\hat{p}^3)^j
	|\hat{\bf X}|^l \hat{P}_l\, (\hat{p}^3)^{k-j}|n\ra
	=  \sum\limits_{j=0}^k\binomial{k}{j}
	   \la n|(\gamma^0\gamma^3)^k(\hat{p}^3)^j
	   |\hat{\bf X}|^l \hat{P}_l\, (\hat{p}^3)^{k-j}|n\ra 
	\;.\label{App:sym-Hu+Hd-2} 
\ee
Applying Eq.~(\ref{App:sym-2}) to the intermediate result 
(\ref{App:sym-Hu+Hd-2}) yields
\be
	\sum\limits_{j=0}^k\binomial{k}{j}
	\la n| \,(\gamma^0\gamma^3)^k\,(\hat{p}^3)^{j}
	|\hat{\bf X}|^l\hat{P}_l\,(\hat{p}^3)^{k-j}|n\ra 
	= (-1)^l \sum\limits_{j=0}^k\binomial{k}{j}
	  \la n| \,(\gamma^0\gamma^3)^k\,(\hat{p}^3)^{j}
	  |\hat{\bf X}|^l\hat{P}_l\,(\hat{p}^3)^{k-j}|n\ra 
	\;,\label{App:sym-Hu+Hd-3} \ee
i.e. the expression vanishes if $l$ is odd. 
The result (\ref{App:sym-Hu+Hd-3}) yields 
the expression in Eq.~(\ref{Hu+Hd-model-mom2}).

\paragraph*{(ii) $(E^u-E^d)$.} 
Here $\Gamma=(1+\gamma^0\gamma^3)\tau^a$ with
$\tilde{\Gamma}=(-1+\gamma^0\gamma^3)\tau^a$. 
It is convenient to take $F(\hat{\bf X})$ as in Eq.~(\ref{App:an-cont-II-7}), 
$F(\hat{\bf X})=$
$(\frac{\partial\;}{\partial\hat{X}^b}|\hat{\bf X}|^l\hat{P}_l)$
$\equiv(\hat{\nabla}^b|\hat{\bf X}|^l\hat{P}_l)$.
Note that the derivative acts on $|\hat{\bf X}|^l\hat{P}_l$ only.
The symmetry transformations (\ref{App:sym-1},~\ref{App:sym-2}) yield 
\ba
&&	\sum\limits_{j=0}^k\binomial{k}{j}
	\la n|(\gamma^0\gamma^3)^{k+1}\tau^a\,(\hat{p}^3)^{j} 
	(\hat{\nabla}^b|\hat{\bf X}|^l\hat{P}_l)(\hat{p}^3)^{k-j}|n\ra
	\nonumber\\
&&	= (-1)^l\sum\limits_{j=0}^k\binomial{k}{j} 
	\la n|(\gamma^0\gamma^3)^{k+1}\tau^a\,(\hat{p}^3)^{j} 
	(\hat{\nabla}^b|\hat{\bf X}|^l\hat{P}_l)(\hat{p}^3)^{k-j}|n\ra
	\;, \label{App:sym-Eu-Ed-1}\ea
i.e. also here contributions of odd $l$ vanish. 
Applying Eq.~(\ref{App:sym-Eu-Ed-1}) to $M^{(m)}_E(\xi,0)$ and using the 
identity $\hat{\nabla}^b F(\hat{\bf X})=i[\hat{p}^b,F(\hat{\bf X})]$ 
we obtain the result quoted in Eq.~(\ref{Eu-Ed-model-mom2}).

\section{Irreducible tensor operators}
\label{App:irreducible-tensor-op}

The grand-spin operator is defined as
\be\label{App:grand-spin}
	\hat{\bf K} = \hat{\bf L}+\hat{\bf S}+\hat{\bf T}\;,
\ee
where $\hat{\bf L}=\hat{\bf r}\times\hat{\bf p}$ is the
orbital angular momentum operator, 
$\hat{\bf S} = \frac12\gamma_5\gamma^0\bgam$
is the spin operator and $\hat{\bf T}=\frac12\btau$ is the isospin operator.
$\hat{\bf K}$ is the generator of simultaneous rotations $R$ in space and 
$SU(2)$-flavour space. Let ${\bf n}$ be a unit vector defining an axis in 
space and flavour space and let $\alpha$ be an angle. 
The rotation $R$ is then characterized by the unitary operator $\hat{U}(R)$
\be\label{App:irr-tens-U}
	\hat{U}(R) = \hat{U}({\bf n},\alpha) 
	           = \exp(-i\,\alpha\,{\bf n}\cdot\hat{\bf K}) \;.\ee
The unitary transformation in Eq.~(\ref{App:irr-tens-U}) transforms
single quark states into ``rotated quark states'' according to 
\[
	\hat{U}(R)|E_n,\pi,K, M\ra = \sum\limits_{M'=-K}^K 
	|E_n,\pi,K, M'\ra D^{(K)}_{M'M}(R)\;.
\]
Here $D^{(K)}_{M'M}(R)$ denote finite-rotation Wigner matrices.

The operators $\hat{T}^{L}_M$ with integers $L$, $M\in [-L,\,L]$ are said to
be irreducible tensor operators of rank $L$, if they transform into each other
under the action of the unitary transformation in Eq.~(\ref{App:irr-tens-U}) 
according to 
\be\label{App:irr-tens-T1}
	\hat{U}(R) \,\hat{T}^{L}_M\, \hat{U}^\dag(R) = 
	\sum\limits_{M' = -L}^L  D^{(L)}_{M'M}(R)\;\hat{T}^{L}_{M'} \; .
\ee
By considering infinitesimal rotations one arrives at a definition,
which is equivalent to Eq.~(\ref{App:irr-tens-T1}),
\ba
	\left[\,\hat{K}^3\,,\;\hat{T}^{L}_M\,\right]  
	&=&	M\,\hat{T}^{L}_M \; , \nonumber\\
	\left[\,\hat{K}^\pm ,\,\hat{T}^{L}_M\,\right]
	&=&	\sqrt{L(L+1)-M(M\pm 1)\,} \;\hat{T}^{L}_{M\pm1} 
	\;,\;\;\mbox{where} \;\;\;
	\hat{K}^\pm \equiv \hat{K}^1 \pm i\hat{K}^2 \;,
	\label{App:irr-tens-T2} \ea
but allows more easily to check whether a given operator is an irreducible 
tensor operator or not. The Hamiltonian $\hat{H}_{\rm eff}$ transforms as 
$\hat{U}(R)\hat{H}_{\rm eff}\hat{U}^{-1}(R)=\hat{H}_{\rm eff}$, since 
it commutes with $\hat{\bf K}$. This means that $\hat{H}_{\rm eff}$ is 
an irreducible tensor operator of rank zero.

Further examples of rank zero operators are $|\hat{\bf X}|$, 
$\btau\cdot\hat{\bf X}$ or $\gamma^0\bgam\cdot\hat{\bf p}$,   
which satisfy Eqs.~(\ref{App:irr-tens-T2}) for $L=M=0$.
The operators
\ba
   \bigl[\hat{T}_a\bigr]^{\,1}_M &=& \cases{
	i(\mp\hat{p}^1-i\hat{p}^2)/\sqrt{2} & for $M=\pm 1$\cr
	i\hat{p}^3                          & for $M=0$,}\nonumber\\
   \bigl[\hat{T}_b\bigr]^{\,1}_M &=& \cases{
	i(\mp\tau^1-i\tau^2)/\sqrt{2} & for $M=\pm 1$\cr
	i\tau^3                       & for $M=0$,}\nonumber\\ 
   \bigl[\hat{T}_c\bigr]^{\,1}_M &=& \cases{
	i(\mp\hat{X}^1-i\hat{X}^2)/\sqrt{2} & for $M=\pm 1$\cr
	i\hat{X}^3                          & for $M=0$,}\nonumber\\
   \bigl[\hat{T}_d\bigr]^{\,1}_M &=& \cases{
	i(\mp\gamma^0\gamma^1
	-i\gamma^0\gamma^2)/\sqrt{2} & for $M=\pm1$\cr
	i\gamma^0\gamma^3            & for $M=0$,}\nonumber\\
   \bigl[\hat{T}_e\bigr]^{\,1}_M &=& \cases{
	i(\mp(\btau\times\hat{\bf p})^1-
	i(\btau\times\hat{\bf p})^2)/\sqrt{2} & for $M=\pm 1$\cr
	i(\btau\times\hat{\bf p})^3           & for $M=0$} 
	\label{App:irr-tens-gamma}\ea
are irreducible tensor operators of rank 1 since they satisfy 
Eqs.~(\ref{App:irr-tens-T2}) for $L=1$. 

Finally Legendre polynomials $P_l(\cos\hat{\theta})$ are irreducible 
spherical tensor operators $\hat{T}^l_m$ with $m=0$, since 
$P_l(\cos\hat{\theta})\propto Y^l_m(\hat{\Omega})|_{m=0}
\equiv\hat{T}^l_m|_{m=0}$. The spherical harmonics $Y^l_m(\hat{\Omega})$ 
are irreducible tensor operators of rank $l$,
since they satisfy Eqs.~(\ref{App:irr-tens-T2}).



\begin{thebibliography}{99} 

\bibitem{Muller:1998fv}
   D.~M\"uller, D.~Robaschik, B.~Geyer, F.~M.~Dittes and J.~Ho\u{r}ej\u{s}i,
   Fortsch.\ Phys.\  {\bf 42}, 101 (1994). 
\bibitem{Radyushkin:1996ru}
   A.~V.~Radyushkin, Phys.\ Lett.\ B {\bf 385}, 333 (1996); 
   A.~V.~Radyushkin, Phys.\ Rev.\ D {\bf 56}, 5524 (1997). 
\bibitem{Ji:1996ek}
   X.~D.~Ji, Phys.\ Rev.\ Lett.\  {\bf 78}, 610 (1997); 
   X.~D.~Ji, Phys.\ Rev.\ D {\bf 55}, 7114 (1997). 
\bibitem{Collins:1996fb}
   J.~C.~Collins, L.~Frankfurt and M.~Strikman,
   Phys.\ Rev.\ D {\bf 56}, 2982 (1997). 
\bibitem{Ji:1998pc}
   X.~D.~Ji, J.\ Phys.\ G {\bf 24}, 1181 (1998). 
\bibitem{Radyushkin:2000uy}
   A.~V.~Radyushkin,
   in {\sl At the frontier of particle physics}, ed. M.~Shifman 
   (World Scientific, Singapore, 2001), vol.~2, p.~1037. 
\bibitem{Goeke:2001tz}
   K.~Goeke, M.~V.~Polyakov and M.~Vanderhaeghen,
   Prog.\ Part.\ Nucl.\ Phys.\  {\bf 47}, 401 (2001). 
\bibitem{factorization-DVCS}
   X.~D.~Ji and J.~Osborne, Phys.\ Rev.\ D {\bf 58}, 094018 (1998);\\ 
   A.~V.~Radyushkin, Phys.\ Rev.\ D {\bf 58}, 114008 (1998);\\ 
   J.~C.~Collins and A.~Freund, Phys.\ Rev.\ D {\bf 59}, 074009 (1999).
\bibitem{Airapetian:2001yk}
   A.~Airapetian {\it et al.}  [HERMES Collaboration],
   Phys.\ Rev.\ Lett.\  {\bf 87}, 182001 (2001).
\bibitem{Stepanyan:2001sm}
   S.~Stepanyan {\it et al.}  [CLAS Collaboration],
   Phys.\ Rev.\ Lett.\  {\bf 87}, 182002 (2001).
\bibitem{Saull:1999kt}
   P.~R.~Saull  [ZEUS Collaboration], 
   in {\sl Proceedings of the International Europhysics
   Conference on High Energy Physics EPS-HEP 99},
   eds. K.~Huitu, H.~Kurki-Suonio, J.~Maalampi
   (IOP, Bristol, UK, 2000), p.~420, hep-ex/0003030.
\bibitem{Favart:2001yj}
   L.~Favart  [H1 Collaboration],
   in {\sl Proceedings of 9th International Conference DIS 2001},
   eds. G.~Bruni, G.~Iacobucci and R.~Nania
   (World Scientific, Singapore, 2002) in press, hep-ex/0106067;
   C.~Adloff {\it et al.}  [H1 Collaboration],
   Phys.\ Lett.\ B {\bf 517}, 47 (2001).
\bibitem{Ji:1997gm}
   X.~D.~Ji, W.~Melnitchouk and X.~Song, Phys.\ Rev.\ D {\bf 56}, 5511 (1997).
\bibitem{Petrov:1998kf}
   V.~Y.~Petrov, P.~V.~Pobylitsa, M.~V.~Polyakov, I.~B\"ornig, K.~Goeke and 
   C.~Weiss, Phys.\ Rev.\ D {\bf 57}, 4325 (1998).
\bibitem{Penttinen:1999th}
   M.~Penttinen, M.~V.~Polyakov and K.~Goeke,
   Phys.\ Rev.\ D {\bf 62}, 014024 (2000).
\bibitem{Christov:1995hr}
   C.~V.~Christov, A.~Z.~G\'orski, K.~Goeke and P.~V.~Pobylitsa,
   Nucl.\ Phys.\ A {\bf 592}, 513 (1995).
\bibitem{Christov:1995vm} 
   C.~V.~Christov {\it et al.},
   Prog.\ Part.\ Nucl.\ Phys.\  {\bf 37}, 91 (1996).
\bibitem{Diakonov:1996sr}
   D.~I.~Diakonov, V.~Y.~Petrov, P.~V.~Pobylitsa, M.~V.~Polyakov and C.~Weiss,
   Nucl.\ Phys.\ B {\bf 480}, 341 (1996);\\
   P.~V.~Pobylitsa and M.~V.~Polyakov, Phys.\ Lett.\ B {\bf 389}, 350 (1996).
\bibitem{Diakonov:1997vc}
   D.~I.~Diakonov {\it et al.}, Phys.\ Rev.\ D {\bf 56}, 4069 (1997).
\bibitem{Pobylitsa:1998tk}
   P.~V.~Pobylitsa {\it et al.}, Phys.\ Rev.\ D {\bf 59}, 034024 (1999);\\
   M.~Wakamatsu and T.~Kubota, Phys.\ Rev.\ D {\bf 60}, 034020 (1999);\\
   K.~Goeke {\it et al.}, Acta Phys.\ Polon.\ B {\bf 32}, 1201 (2001);\\
   P.~Schweitzer {\it et al.}, Phys.\ Rev.\ D {\bf 64}, 034013 (2001).
\bibitem{Polyakov:1999gs}
   M.~V.~Polyakov and C.~Weiss, Phys.\ Rev.\ D {\bf 60}, 114017 (1999).
\bibitem{Belitsky:2000vk}
   A.~V.~Belitsky, D.~M\"uller, A.~Kirchner and A.~Sch\"afer,
   Phys.\ Rev.\ D {\bf 64}, 116002 (2001).
\bibitem{Teryaev:2001qm}
   O.~V.~Teryaev, Phys.\ Lett.\ B {\bf 510}, 125 (2001).
\bibitem{Kivel:2000fg}
   N.~Kivel, M.~V.~Polyakov and M.~Vanderhaeghen,
   Phys.\ Rev.\ D {\bf 63}, 114014 (2001).
\bibitem{Diakonov:1987ty}
   D.~I.~Diakonov, V.~Y.~Petrov and P.~V.~Pobylitsa,
   Nucl.\ Phys.\ B {\bf 306}, 809 (1988); \\
   D.~I.~Diakonov and V.~Y.~Petrov, JETP Lett.\  {\bf 43}, 75 (1986)
   [Pisma Zh.\ Eksp.\ Teor.\ Fiz.\  {\bf 43}, 57 (1986)].
\bibitem{Diakonov:tw}
   D.~I.~Diakonov and M.~I.~Eides, JETP Lett.\  {\bf 38}, 433 (1983)
   [Pisma Zh.\ Eksp.\ Teor.\ Fiz.\  {\bf 38}, 358 (1983)].
\bibitem{Dhar:gh}
   A.~Dhar, R.~Shankar and S.~R.~Wadia,	Phys.\ Rev.\ D {\bf 31}, 3256 (1985). 
\bibitem{Diakonov:1985eg}
   D.~I.~Diakonov and V.~Y.~Petrov, Nucl.\ Phys.\ B {\bf 272}, 457 (1986).
\bibitem{Diakonov:1983hh}
   D.~I.~Diakonov and V.~Y.~Petrov, Nucl.\ Phys.\ B {\bf 245}, 259 (1984).
\bibitem{Witten:tx}
   E.~Witten, Nucl.\ Phys.\ B {\bf 223}, 433 (1983).
\bibitem{Gluck:1994uf}
   M.~Gl\"uck, E.~Reya and A.~Vogt, Z.\ Phys.\ C {\bf 67}, 433 (1995);\\
   M.~Gl\"uck, E.~Reya, M.~Stratmann and W.~Vogelsang,
   Phys.\ Rev.\ D {\bf 53}, 4775 (1996).
\bibitem{Fano-Racah}
   U.~Fano and G.~Racah, {\sl Irreducible tensorial sets}
   (Academic Press, New York, 1959), pp.~79.
\bibitem{Efremov:1979qk}
   A.~V.~Efremov and A.~V.~Radyushkin, Phys.\ Lett.\ B {\bf 94}, 245 (1980).
\bibitem{Lepage:1979zb}
   G.~P.~Lepage and S.~J.~Brodsky, Phys.\ Lett.\ B {\bf 87}, 359 (1979).
\bibitem{Vanderhaeghen:1999xj}
   M.~Vanderhaeghen, P.~A.~M.~Guichon and M.~Guidal,
   Phys.\ Rev.\ D {\bf 60}, 094017 (1999).
\bibitem{Korotkov:2001zn}
   V.~A.~Korotkov and W.~D.~Nowak, Eur.\ Phys.\ J.\ C {\bf 23}, 455 (2002).
\bibitem{Belitsky:2001ns}
   A.~V.~Belitsky, D.~M\"uller and A.~Kirchner,
   Nucl.\ Phys.\ B {\bf 629} 323 (2002).
\bibitem{Polyakov-private-communication}
   M.~V.~Polyakov, private communication.
\bibitem{Weinberg+Wigner}
   S.~Weinberg, ``The Quantum Theory of Fields. Vol.~1: Foundations''
   (Cambridge University Press, Cambridge, UK, 1995) p.~100.
   The original idea is due to 
   E.~P.~Wigner, in ``Group Theoretical Concepts and Methods in Elementary
   Particle Physiscs'', ed. F.~G\"ursay (Gordon and Beach, New York, 1964) 
   p.~37.
\bibitem{Anselmino:2002yx}
   M.~Anselmino, V.~Barone, A.~Drago and F.~Murgia,
   {\tt hep-ph/0209073}.
\bibitem{Magnus+Oberhettinger}
   W.~Magnus and F.~Oberhettinger,
   {\sl Formulas and Theorems for the Special Functions of Mathematical 
   Physics} (Chelsea Publishing Company, New York, 1949), pp.~50.

\end{thebibliography}
\end{document}